\documentclass[]{aa}
\usepackage{graphicx}
\usepackage[varg]{txfonts}
\usepackage{hyperref}
\hypersetup{colorlinks, citecolor=black, linkcolor=black}
\usepackage{amsmath}

\begin{document}
\title{Deriving structural parameters of semi-resolved star clusters}
\subtitle{FitClust: a program for crowded fields}

\author{D. Narbutis\inst{1} \and D. Semionov\inst{1} \and R. Stonkut{\. e}\inst{1} \and P. de Meulenaer\inst{1,2} \and T. Mineikis\inst{1,2} \and A. Brid{\v z}ius\inst{1,2} \and V. Vansevi{\v c}ius\inst{1,2}}

\institute{Center for Physical Sciences and Technology, Savanori{\c u} 231, LT-02300 Vilnius, Lithuania \\\email{donatas.narbutis@ftmc.lt} \and Vilnius University Observatory, {\v C}iurlionio 29, LT-03100 Vilnius, Lithuania}

\date{Received 30 August 2013, Accepted 19 June 2014}

\abstract
{An automatic tool to derive structural parameters of semi-resolved star clusters located in crowded stellar fields in nearby galaxies is needed for homogeneous processing of archival frames.}
{We have developed a program that automatically derives the structural parameters of star clusters and estimates errors by accounting for individual stars and variable sky background.}
{Models of observed frames consist of the cluster's surface brightness distribution, convolved with a point spread function; the stars, represented by the same point spread function; and a smoothly variable sky background. The cluster's model is fitted within a large radius by using the Levenberg-Marquardt and Markov chain Monte Carlo algorithms to derive structural parameters, the flux of the cluster, and individual fluxes of all well-resolved stars.}
{{\sc FitClust}, a program to derive structural parameters of semi-resolved clusters in crowded stellar fields, was developed and is available for free use. The program was tested on simulated cluster frames, and was used to measure clusters of the M31 galaxy in Subaru Suprime-Cam frames.}
{Accounting for bright resolved stars and variable sky background significantly improves the accuracy of derived structural parameters of star clusters. However, their uncertainty remains dominated by the stochastic noise of unresolved stars.}


\titlerunning{Structural parameters of star clusters}
\authorrunning{D. Narbutis et al.}

\maketitle

\section{Introduction}
An automatic tool to derive structural parameters of star clusters is essential so that numerous archival frames from stellar population surveys in nearby galaxies can be processed homogeneously. Observed surface brightness distribution in frames is commonly treated as a sum of the sky background and the cluster's profile, e.g., in Hubble Space Telescope (HST) studies of clusters in galaxies: M33 \citep{SanRoman2012}, M31 \citep{Tanvir2012}, and NGC\,7252 \citep{Bastian2013}, or ground based observations of M31 clusters \citep{Barmby2007}. However, the aforementioned studies are focused mostly on bright clusters. For fainter clusters, located in crowded fields of galaxy disks, the brightest individual stars have to be taken into account properly in order to avoid bias in structural parameters, derived by fitting cluster models to the surface brightness distribution.

We propose a method to solve an inverse problem aimed at finding best-fit for 1) a cluster; 2) an unknown number of bright stars (cluster members or foreground objects); and 3) a variable sky background.

Structural parameters of clusters are degenerate and are due to problems of a sky background determination in crowded areas \citep{Werchan2011}. It is virtually impossible to determine a true sky background in a crowded field frame because of unresolved sources of variable spatial number density. For some clusters in the M31 galaxy \citep{Narbutis2008}, which reside on dust lanes, severe background variability was observed. The strongest effect due to the sky background determination inaccuracy is on the parameters describing luminosity distribution in the outskirts of clusters, e.g., tidal radius.

In a study of a semi-resolved star cluster sample compiled by \cite{Kodaira2004} from a Subaru Suprime-Cam survey of the M31 galaxy, a grid of model surface brightness profiles convolved with the point spread function (PSF) was constructed and used to derive parameters of the observed clusters \citep{Sableviciute2007}. However, this method did not take into account individual stars and variability of the sky background.

\cite{Larsen1999} developed a program ISHAPE to derive structural parameters of unresolved clusters based on masking out deviating pixels. It was applied to the study of the M31 clusters \citep{Sableviciute2006}. However, for extended objects significantly larger than the size of the PSF, stars could not be masked out successfully.

Frames of crowded fields are usually processed using DAOPHOT \citep{Stetson1987} or similar programs to make stellar photometry catalogs. However, they have a resolution limit, and split unresolved cores of clusters into several artificial point sources. Hence, they are suitable for construction of color-magnitude diagrams of the brightest stars only in the outer parts of semi-resolved clusters, e.g., an extended object studied by \cite{Stonkute2008} or clusters measured in HST frames by \cite{Larsen2011}.

Therefore, a robust tool suitable for deriving cluster parameters in crowded stellar fields is needed.

In this paper we present a program, {\sc FitClust}, designed to derive structural parameters of clusters with overlapping bright stars located on a variable sky background. {\sc FitClust} is implemented in Python and uses DAOPHOT within PyRAF. The code can be obtained from the {\sc FitClust} website\footnote{{\sc FitClust}: www.astro.ff.vu.lt/software/fitclust}.

In Sect. 2 we describe a method implemented in the program; in Sect. 3 we analyze its performance with artificial clusters and apply it to the study of clusters in the M31 galaxy.

\section{Deriving structural parameters}
In order to derive structural cluster parameters by employing {\sc FitClust}, the following input data are required:
\begin{itemize}
\item an observed frame containing a cluster;
\item a normalized PSF (i.e., the sum of pixel values is set to unity) derived for the frame;
\item a bad pixel mask;
\item a model of a cluster as an analytic 2D brightness distribution (coordinates, shape parameters, and total flux);
\item coordinates and fluxes of the resolved stars identified and fitted with the PSF;
\item a sky background: a fixed value (for the flat sky) or an approximation with a 2D spline (for the variable sky);
\item a set of algorithm control parameters.
\end{itemize}

The basic steps of the iterative cluster and star fitting algorithm implemented in {\sc FitClust} are displayed in Fig.\,1. The main loop of model fitting consists of 1) measurement and subtraction of sky background; 2) fit of the cluster and stars together; 3) subtraction of the best-fit cluster to produce a residual frame; 4) derivation of more accurate coordinates of stars on the residual frame; 5) subtraction of the cluster and stars to measure sky background for the next iteration; and 6) sampling of parameter space with Markov chain Monte Carlo (MCMC) algorithm.

\begin{figure}
\includegraphics[width=88mm]{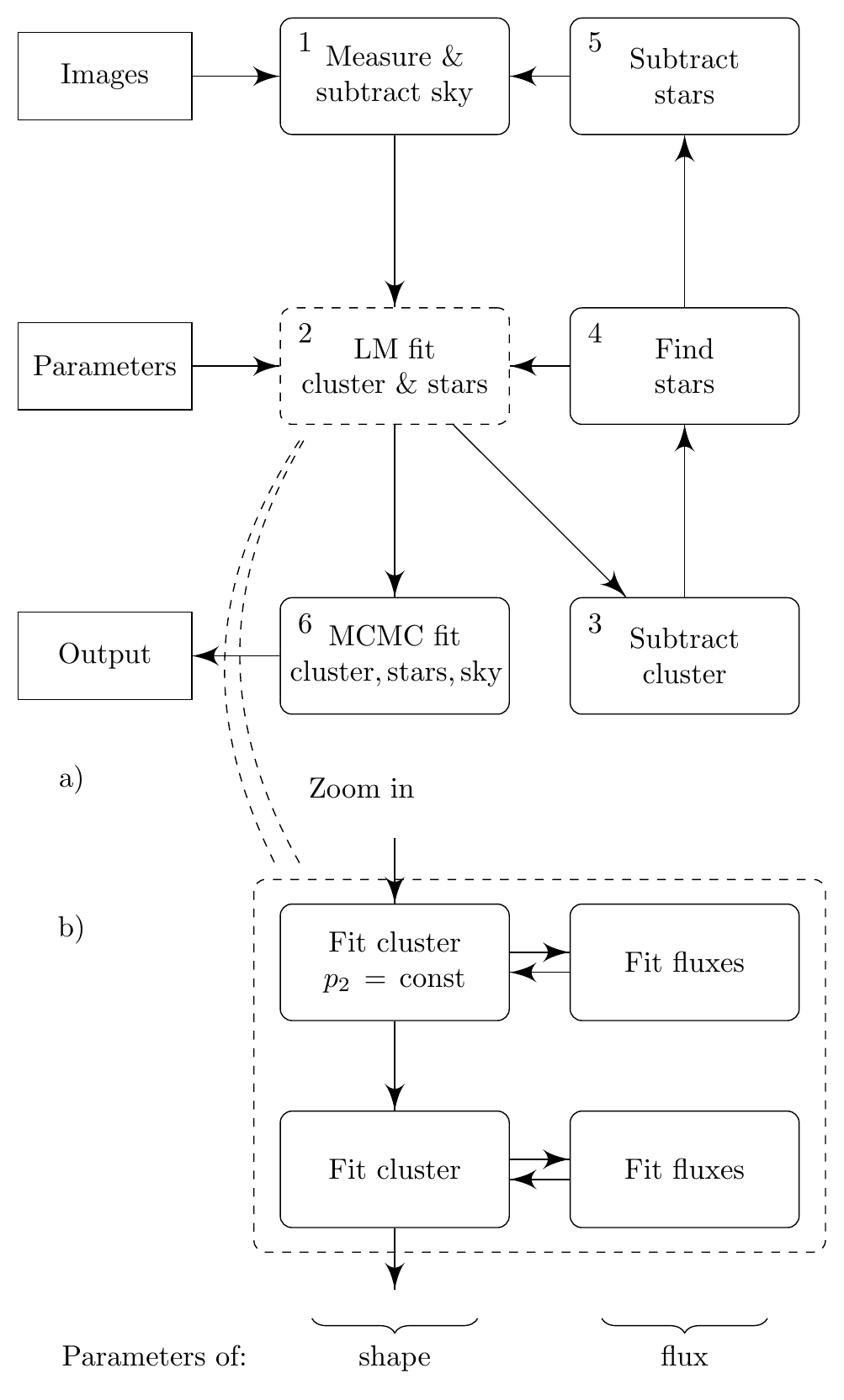}
\caption{Schematic representation of the iterative cluster and star fitting algorithm with {\sc FitClust}. Panel a) shows the basic steps and data blocks, while panel b) elaborates on the innermost loop of cluster parameter fitting, which are separated into two groups.}
\end{figure}

The main loop runs for a given number of iterations and produces the best-fit cluster model parameters, which can optionally be fitted using MCMC algorithm to derive their uncertainties. The parameter values of model components are derived by minimizing an array of residuals within specified model fitting area, weighting each pixel by its uncertainty assuming photon noise.

\subsection{Input frames}
{\sc FitClust} takes a small subframe from the observed frame centered on a cluster and uses it as an input. If the frame contains CCD defects or saturated stars, they are excluded by using a bad pixel mask, which can be created automatically \citep[e.g., with the program StarL by][]{Narbutis2009} and edited manually.

Usually, there are several clusters to be measured in the observed frame. For convenience, {\sc FitClust} can automatically construct the empirical PSF for the entire frame using the DAOPHOT program within the IRAF system \citep{Tody1993}. If the PSF is strongly variable, which is the case for wide-field imaging, the PSF for individual clusters (subframes) can be constructed based on isolated stars located in the cluster's vicinity.

\subsection{Sky background determination}
A sky background value is determined for every iteration (Fig.\,1) of cluster model fitting and subtracted from the input frame. For the first iteration, the original input frame is used to construct a sky value histogram, whose lower pixel value part is fitted with a Gaussian function to derive the initial sky level value. For subsequent iterations, a residual frame with the cluster and stars subtracted is used to measure a sky background either by 1) fitting a sky value with a Gaussian function; or 2) fitting a 2D spline function.

\subsection{Parameters}
The following parameters are used to control program execution: 1) a type of cluster model surface brightness distribution; 2) approximate initial coordinates (identified by eye, by computing mass center of flux distribution using moments, or by fitting a 2D Gaussian function) of a cluster center, $x_{\rm in}, y_{\rm in}$; 3) a radius of a circle defining the fitting area of the model, $r_{\rm fit}$; 4) a number of fitting and star detection iterations, $n_{\rm fit}$; and 5) the limits for fitting parameters.

A projection of an unresolved cluster on the frame is usually described by tidal-cut-off King \citep{King1962} or power-law EFF \citep{Elson1987} profiles, which are used to simulate a smooth surface brightness distribution. A frame of a cluster model is constructed using values of an analytic function at the center of each pixel and normalizing flux to unity. For the profiles of the EFF and King models, a truncation radius is used for normalization.

The King surface brightness profile is defined by a central surface brightness, $\mu_0$; a core radius, $r_c$; and a tidal radius, $r_t$:
\begin{equation}
\mu(r)=\mu_0\left[{\left({1+\frac{{r^2}}{{r_c^2}}}\right)^{-1/2}-\left({1+\frac{{r_t^2}}{{r_c^2}}}\right)^{-1/2}}\right]^2.
\end{equation}
The EFF profile is defined by $\mu_0$; a scale radius, $r_e$; and a power-law index, $n$:
\begin{equation}
\mu(r)=\mu_0\left({1+\frac{{r^2}}{{r_e^2}}}\right)^{-n}.
\end{equation}

A cluster model is described by center coordinates, $x_c, y_c$; two constants defining a shape of the profile, $p_1, p_2$; and a total flux, $f_c$, which is linked to the central surface brightness, $\mu_0$. Analytic brightness distribution is convolved using FFTW library\footnote{http://www.fftw.org} with the PSF to mimic observation effects. Stars are represented by the PSF shifting it to the desired positions, $x_i, y_i$, using spline interpolation and scaling by fluxes, $f_i$.

Therefore, the following cluster model parameters are provided:
\begin{itemize}
\item $x_c, y_c$ -- center coordinates of a cluster;
\item $p_1, p_2$ -- for the King model: the core radius, $r_c$, and the ratio of tidal to core radii, $r_t/r_c$; for the EFF model: the scale radius, $r_e$, and the power-law index, $n$;
\item $f_c$ -- the total flux of a cluster;
\item $x_i, y_i, f_i$ -- center coordinates and the flux of an $i$-th star;
\item $\mu_{\rm sky}$ -- the sky background: constant or approximated by a 2D spline function.
\end{itemize}

\subsection{Model fitting}
The loop of model fitting shown in Fig.\,1 can be summarized as a following sequence of steps repeated for a given number of iterations, $n_{\rm fit}$:
\begin{itemize}
\item measure sky background, $\mu_{\rm sky}$, on 1) an input frame in the first iteration; 2) a residual frame (with a cluster and stars subtracted) in subsequent iterations; subtract sky background from the input frame;
\item derive cluster parameters $x_c, y_c, p_1, p_2$ by 1) fitting only a cluster in the first iteration; 2) fitting cluster and stars in the subsequent iterations while keeping their coordinates, $x_i, y_i$, fixed (derived during previous iteration); derive flux of a cluster, $f_c$, and of stars, $f_i$, by performing algorithm steps defined in Fig.\,1\,(b);
\item make a residual frame by subtracting fitted cluster model from the input frame;
\item use DAOPHOT on the residual frame to find resolved stars and measure their coordinates, $x_i, y_i$; pass list of star coordinates to the next iteration of model fitting;
\item subtract stars from the residual frame and pass it to the next iteration of sky background determination.
\end{itemize}

To derive parameters of a cluster as shown in Fig.\,1\,(b), we use the Levenberg-Marquardt (LM) algorithm\footnote{http://cars9.uchicago.edu/software/python/lmfit} and constrain parameter limits. We have found that fitting cluster coordinates, $x_c, y_c$, and shape parameters, $p_1, p_2$, simultaneously does not perform correctly, because solutions get stuck in local minima. Therefore, we run the first pass of LM algorithm and fit only $x_c, y_c, p_1$, while $p_2 = {\rm const}$ (King $r_t/r_c = 10$ and EFF $n = 1.5$). The coordinates and both shape parameters are fitted in the second pass of LM algorithm.

The model frame is constructed from a cluster model and a reasonable number of stars ($N \lesssim 20$, because time of fitting is proportional to $N^2$), all stored as individual 2D arrays with volumes normalized to unity. For each pixel of a residual frame we have
\begin{equation}
\mu_{\rm res}=\mu_{\rm in} - \mu_{\rm sky} - \mu_{\rm mod},
\end{equation}
\begin{equation}
\mu_{\rm mod}=f_c \cdot \mu_c + \sum_{i=1}^N f_i \cdot \mu_{\rm PSF}.
\end{equation}
Here $\mu_{\rm res}$ is a pixel value of the residual frame; $\mu_{\rm in}$ of the input frame; $\mu_{\rm sky}$ of the sky background; $\mu_{\rm mod}$ of the model frame, which contains a cluster flux, $f_c$, multiplied by cluster model, $\mu_c$; and a sum of star fluxes, $f_i$, multiplied by the PSF model, $\mu_{\rm PSF}$. For the LM fitting the fluxes of cluster and stars are constrained to be positive. Ideally, one would like to obtain residuals comparable to the measured sky background noise. Practically, stochastic noise of unresolved stars and variable crowding conditions within the cluster make residuals larger and dependent on the cluster's age, mass, and size.

\subsection{Iterative detection of stars}
Approximate coordinates of a cluster in the frame, $x_{\rm in}, y_{\rm in}$, are used to initialize the fitting procedure. Initially, we fit and subtract only a model of a cluster and pass the residual frame to DAOPHOT.

To find stars on the residual frame a {\tt daofind} routine is used with a detection threshold of 4 standard deviations above the sky background level. Stars are measured with {\tt allstar} and then subtracted. A second pass of {\tt daofind} is performed to find additional stars in the residual frame. Lists of stars are concatenated and a second pass of {\tt allstar} produces a list of star coordinates, $x_i, y_i$. Stars located within a fitting radius, $r_{\rm fit}$, are selected and provided to the next iteration of the model fitting loop. Figure 2 shows a cluster frame decomposed into a cluster model, stars, and a residual frame.

\begin{figure}
\includegraphics[width=88mm]{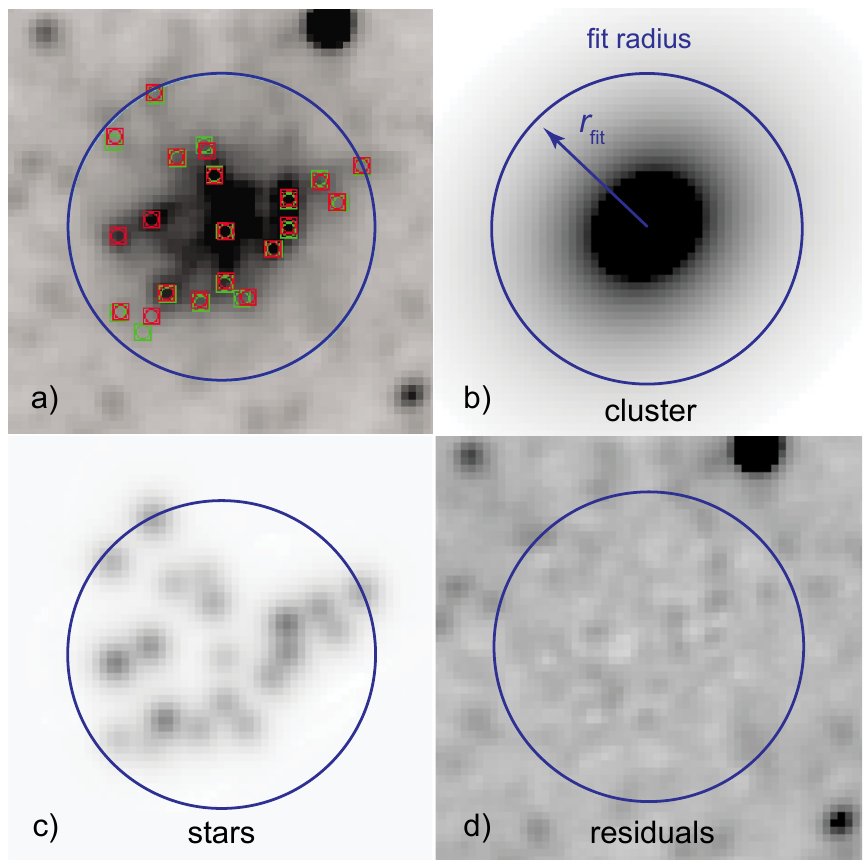}
\caption{Frame of one of the largest clusters from the M31 sample \citep{Vansevicius2009} shown in panel (a); it is decomposed into cluster model (b), stars (c), and residual frame (d). In panel (a) coordinates of stars detected in the first iteration are shown in green, the final iteration in red.}
\end{figure}

In each iteration stars are identified anew and their coordinates are used only for subsequent iteration. Therefore, the number of stars, $N$, and their coordinates, $x_i, y_i$, vary and converge to the best-fit values. Cluster parameters and sky background also converge to the best-fit values, while the standard deviation of residuals, $\sigma_{\rm fit}$, within $r_{\rm fit}$ decreases. The model fitting ends after a preselected number, $n_{\rm fit}$, of iterations -- typically $n_{\rm fit}$ is approximately 6.

\subsection{Parameter fitting with MCMC}
We have found that iterative model fitting performs well when detecting the brightest stars and deriving the parameters of old and massive clusters with smooth profiles. When young clusters of low mass are analyzed, fitting can get stuck in local minima because of stochastic variations of flux in the cluster's profile and stellar crowding.

To avoid local minima and to estimate parameter uncertainties, we use {\tt emcee}, a Python implementation \citep{Mackey2013} of the affine-invariant MCMC ensemble sampler proposed by \cite{Goodman2010}. It starts an ensemble of random walkers in the model parameter space from a position of the best-fit values of cluster, star, and sky background parameters obtained by the LM algorithm.

As each walker makes steps through the parameter space, a model of a cluster with a fixed number of stars is constructed at each step. We note that all cluster parameters, including star positions and fluxes, are allowed to vary when making MCMC steps. At each step a likelihood is computed for the input frame to be generated by the model with given parameter values, assuming that uncertainty of each pixel is Gaussian with a standard deviation equal to the square root of the input pixel's value.

After the burn-in phase, which takes $\sim$$3\,000$ steps when fitting a cluster without stars, {\tt emcee} converges to the most likely parameter values and subsequent walker steps are confined within the limits of parameter uncertainties. The second sample of $\sim$$1\,000$ walker positions is analyzed. The median of the sample and its standard deviation are reported as the most likely value of parameter and its uncertainty.

For a cluster with one star included, $\sim$$6\,000$ steps are required for burn-in. The time needed to create a model image steeply increases with number of included stars. Therefore, practical considerations control the limit on the number of the brightest stars to include.

\subsection{Output}
The final step of the model fitting is numerical integration to derive a cluster's half-light radius, $r_h$. For this purpose a cluster model is produced with a frame resolution that is ten times higher. On this frame a curve of growth is constructed by integrating flux through circular apertures up to a truncation radius, $r_{\rm max}$, where the surface brightness, $\mu$, of a cluster model reaches 1\% of its central value, $\mu_0$.

In addition to the frames shown in Fig.\,2, {\sc FitClust} also outputs a record file containing parameters of a cluster model and stars, as well as $\sigma_{\rm fit}$, which can be used to analyze the quality of fitting.

\section{Tests and results}
We have used {\sc FitClust} to derive structural parameters of clusters in crowded fields of the M31 disk taken from the cluster sample presented by \cite{Narbutis2008}. Examples of some characteristic objects are shown in Fig.\,3. Parameters for a double cluster (see Fig.\,3\,e) with two peaks in luminosity distribution were derived by fitting two cluster models simultaneously. In all three cases the subtracted cluster and stars produce a clean residual frame with a standard deviation, $\sigma_{\rm fit}$, equal to that of sky background. Initially we attempted to fit elliptical models, but because of dominating stochastic effects in low mass clusters, analysis was limited to circular ones.

\begin{figure}
\includegraphics[width=88mm]{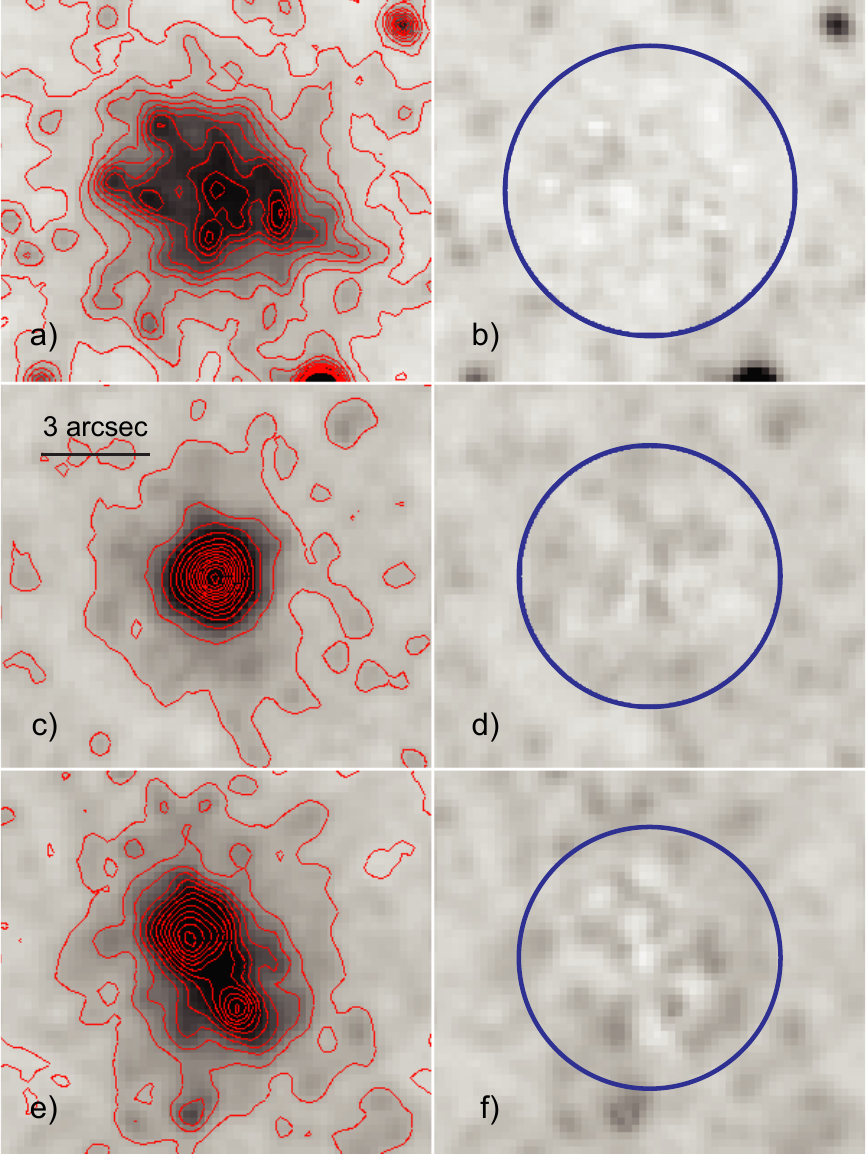}
\caption{{\it Left}: examples of M31 star clusters observed with Suprime-Cam taken from the study by \cite{Vansevicius2009}, over-plotted with iso-flux lines, from top to bottom: semi-resolved, compact, and double. Image scale is indicated in panel (c). {\it Right}: residuals after subtracting the cluster model and stars; circles indicate fitting area.}
\end{figure}

We simulated frames of artificial clusters and analyzed them with {\sc FitClust} in order to evaluate its performance when detecting well-resolved stars automatically, determining sky background, and deriving cluster structural parameters. Artificial cluster frames resemble the quality of those obtained with Suprime-Cam in M31 studied by \cite{Vansevicius2009}: 1) frame scale 0.2 arcsec/pixel; 2) full width at half maximum (FWHM) of the PSF 3 pixels.

In the model fitting routine we set the following limits for cluster structural parameters: $r_c = r_e = [0.01, 15]$ pixels (both models), $r_t/r_c = [2, 500]$ (King), and $n = [1.01, 5]$ (EFF).

Just to illustrate that even in the simplest cases structural parameters are significantly degenerate, we simulated smooth artificial clusters using King and EFF models and convolved them with a PSF. Neither photon noise nor overlapping stars were included in this test. Fitting was performed at the nodes of the grid ($p_2$ vs. $p_1$) with both parameters fixed. The derived maps of fitting standard deviations, $\sigma_{\rm fit}$, are shown in Fig.\,4. A shallow band of $\sigma_{\rm fit}$ minimum is seen over a wide parameter range, where parameters are degenerate. These maps become even more complicated (have few local minima) when several stars are introduced into fitting.

\begin{figure}
\includegraphics[width=88mm]{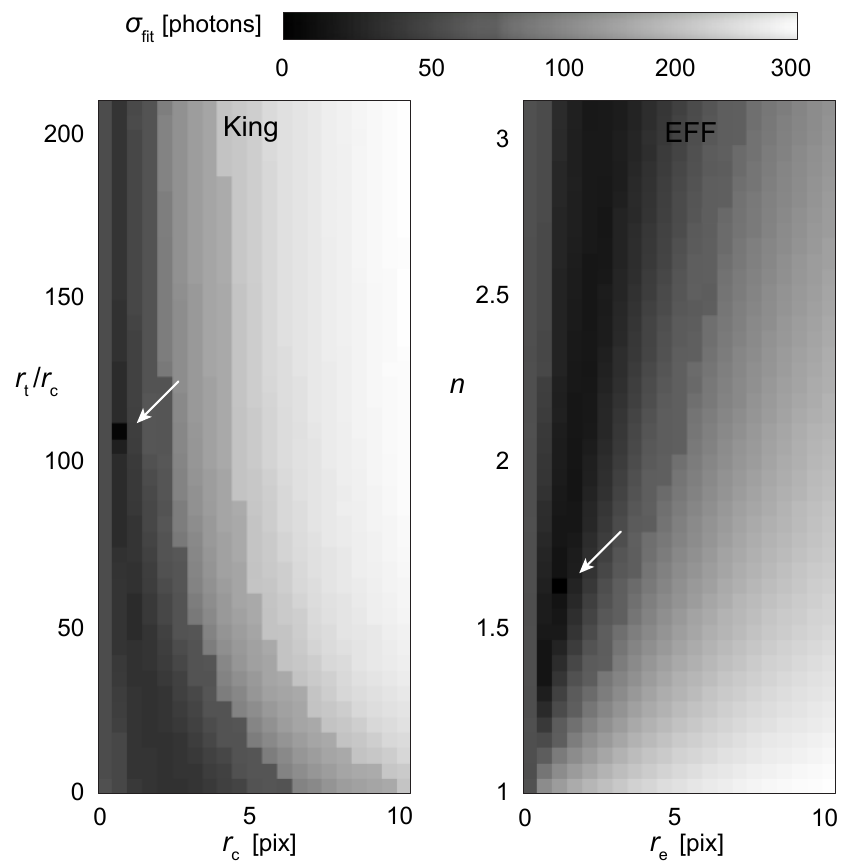}
\caption{Maps of fitting standard deviations, $\sigma_{\rm fit}$, for artificial clusters based on smooth King ({\it left}) and EFF ({\it right}) models. Position of the cluster under consideration in the parameter space is indicated by the arrow ($r_c = 1$ pixel, $r_t/r_c = 110$, and $r_e = 1.5$ pixel, $n = 1.7$) and coincides with minimal value of standard deviation.}
\end{figure}

\subsection{Bright star test}
We tested the algorithm to derive parameters of smooth model cluster in the presence of three resolved stars and uniform sky background, i.e., without faint unresolved objects.

The smooth clusters assuming the King and EFF models were simulated at the nodes of parameter grids displayed as black dots in Figs.\,5 and 6, respectively. Nodes of the parameter grids were chosen to cover the parameter space of M31 clusters with $V < 20$\,mag, taken from the dataset produced for the study by \cite{Vansevicius2009} and are shown with open circles in Figs.\,5\,(c) and 6\,(c). Flux of a model cluster was set to $10^6$ photons, which corresponds to $V \sim 17.5$\,mag for the object in the M31 cluster sample. The cluster model was convolved with a PSF and placed at the center of $101 \times 101$ pixel frame. At each node of the grid series of 100 frames were simulated.

\begin{figure*}
\centering
\includegraphics[width=160mm]{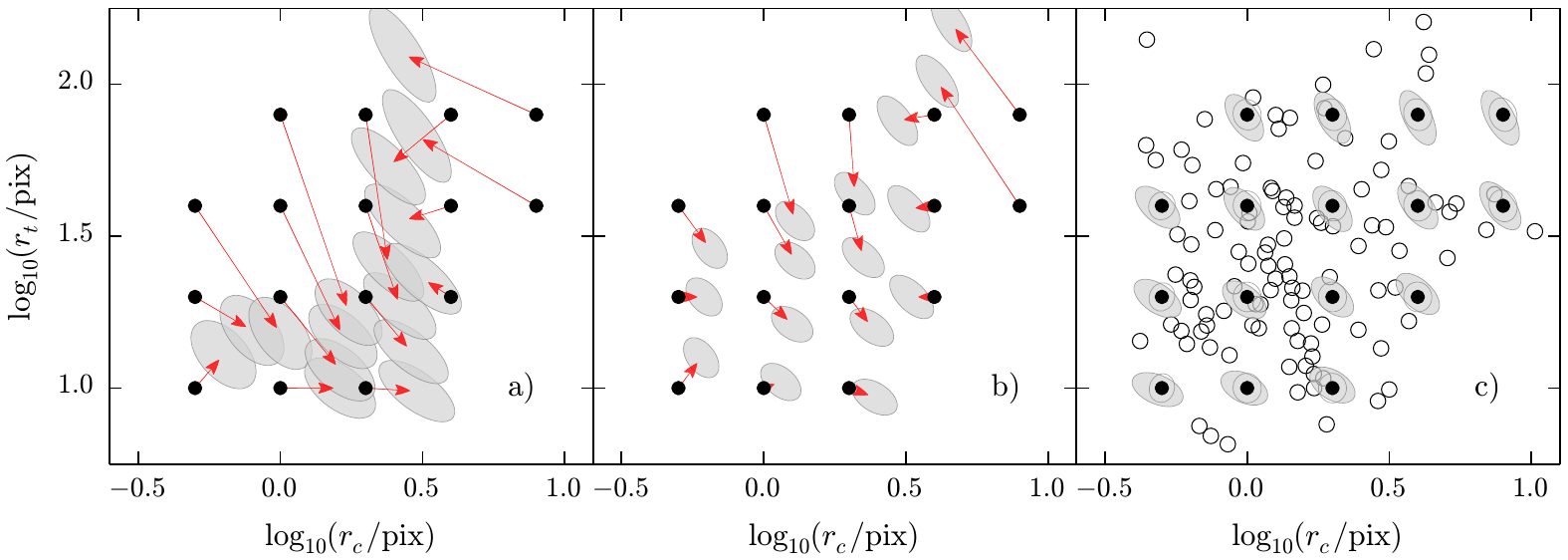}
\caption{Bright star test. Results of the King model fitting on frames of smooth artificial clusters, superposed by three identical stars. The total flux of the cluster is $10^6$ photons. The cases of each star flux of $10^5$ and $4 \times 10^4$ photons are shown in panels (a) and (b), respectively. Panels (a, b) display cases when stars are not included in the model fit, and (c) when they are automatically identified and fitted (smaller ellipses are for fainter stars originally in panel b). Tidal radius, $r_t$, and core radius, $r_c$, are displayed. Dots mark nodes and indicate initial parameters; for each node series of 100 frames are simulated. Ellipses approximate 1--$\sigma$ scatter of the derived parameters and arrows show their bias. Open circles (panel c) show M31 clusters from \cite{Vansevicius2009} with $V < 20$\,mag.}
\end{figure*}

\begin{figure*}
\centering
\includegraphics[width=160mm]{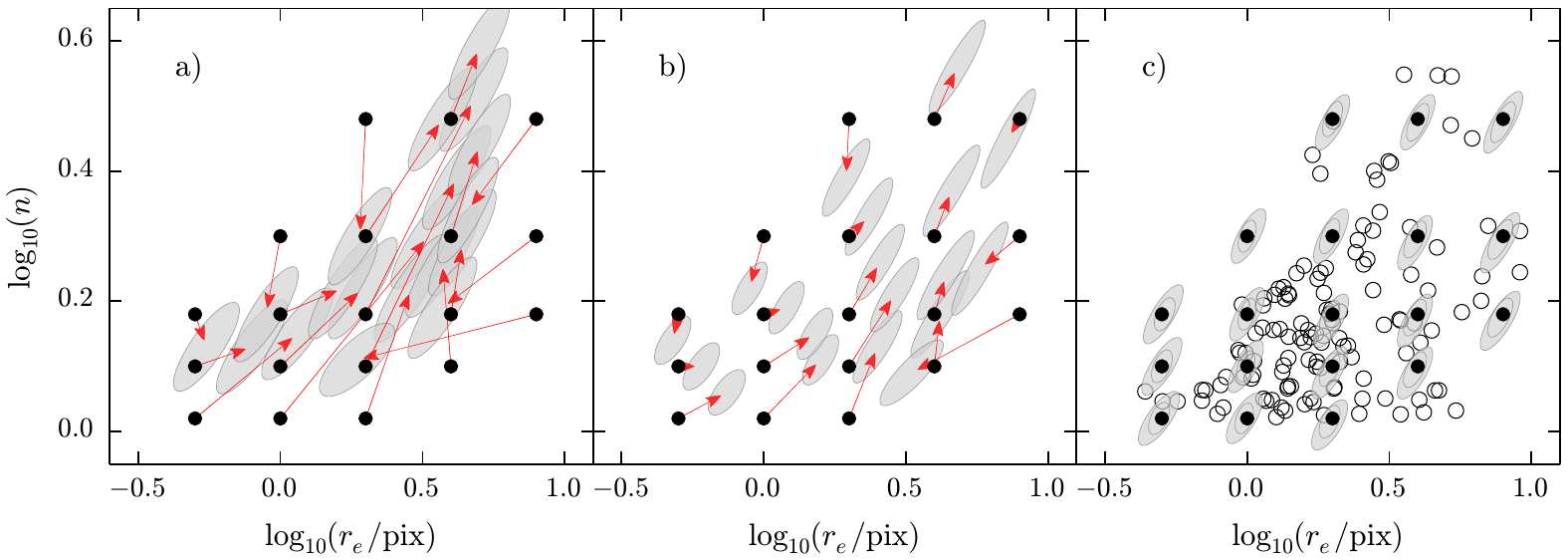}
\caption{Bright star test. The same as Fig.\,5, but for smooth clusters simulated with the EFF model. The power-law index, $n$, and the scale radius, $r_e$ are displayed.}
\end{figure*}

Each cluster was superposed with three stars randomly placed within a circle of a 5 pixel radius centered on the cluster to make star detection more difficult. Figure 5\,(a) shows cases when each interfering star has a flux of $10^5$ photons. Since the total flux of three interfering stars makes up to $\sim$30\% of the cluster's flux, they alter the cluster luminosity distribution significantly.

Uniform sky background was modeled assuming a flux of $10^3$ photons per pixel. The Gaussian photon noise was added to each pixel of the frame, assuming its standard deviation to be equal to the square root of the number of photons in a pixel. The cluster's fitting radius was set to 31 pixels; ten {\sc FitClust} star detection and model fitting iterations were performed.

As can be seen in Fig.\,5\,(a), clusters with a small core radius tend to have larger measured $r_c$, and clusters with large $r_c$ demonstrate an opposite effect. The tidal radius, $r_t$, decreases for all nodes, except for those with the largest $r_c$. Therefore, stars projected on the cluster introduce a significant bias into the derived cluster structural parameters. We note, however, that a similar effect on parameters, as shown in Fig.\,5\,(a), is also seen when even a single star is projected on the cluster. The three stars were chosen to demonstrate {\sc FitClust}'s ability to identify and measure stars projected in various configurations on the cluster.

When automatic detection of stars in {\sc FitClust} is turned on, the derived structural parameters are in a good agreement with the input parameters of simulated clusters, see Fig.\,5\,(c) larger ellipses around grid points. The scatter of the parameters shown by 1--$\sigma$ ellipses is smaller than in the cases when stars are not fitted (Fig.\,5\,a). However, it is non-negligible because of photon noise and limited accuracy of star subtraction procedure. In most cases the parameters of stars are derived correctly. The flux of clusters is derived with an average accuracy of $\sim$5\%.

The results of King model fitting in cases of three stars with flux of $4 \times 10^4$ photons each are shown in Fig.\,5\,(b). The total flux of stars makes $\sim$12\% of the cluster's flux and their disturbance of measured structural parameters is lower. However, directions of systematic shifts remain similar to those seen in Fig.\,5\,(a). Figure 5\,(c, smaller ellipses) shows derived parameters when stars are fitted together with the cluster. The ellipses are smaller than in Fig.\,5\,(b), because the residuals remaining after subtraction of fainter stars are less pronounced. The flux of clusters in these cases is derived with an average accuracy of $\sim$3\%.

The results of cluster parameters derived from frames simulated with the smooth EFF model are shown in Figs.\,6\,(a) and (b) for cases of interfering stars with flux of $10^5$ and $4 \times 10^4$ photons, respectively. As in the case of King model, when stars are not fitted, scale radius, $r_e$, and power-law index, $n$, are significantly biased. However, automatic detection and fit of stars in {\sc FitClust} switched on again gives correct structural cluster parameters (Fig.\,6\,c), and it takes six iterations on average to obtain a consistent solution for stars and the cluster.

\subsection{Sky background test}
To test performance of {\sc FitClust} in realistic sky background conditions, we selected two regions in the M31 Suprime-Cam CCD mosaic ($V$-band) taken from \cite{Narbutis2008}. The first region represents a crowded field with $\sim$20 stars within fitting radius. The second is located on the edge of the prominent dust lane, which introduces a sky background gradient across the $101 \times 101$ pixel frame used to derive structural cluster parameters.

Smooth cluster model frames were generated according to the prescription given in Sect. 3.1, except that they were convolved with the empirical PSF of Suprime-Cam. At each node of the parameter grids 100 cluster frames were generated and overlaid on the subframes of two selected M31 regions by randomly shifting coordinates within a box of 50 pixels. Model fitting radius was set to 25 pixels and seven star detection and model fitting iterations were performed.

To estimate the accuracy of the derived cluster parameters versus cluster brightness, we have repeated the same test with artificial clusters having fluxes of $10^6$, $4 \times 10^5$, and $1.6 \times 10^5$ photons, which correspond to approximate $V$-band magnitudes of 17.5, 18.5, and 19.5\,mag (approximate mass range from $30\,000$\,${\rm M}_\odot$ to $3\,000$\,${\rm M}_\odot$ for the clusters of 100\,Myr age) in Suprime-Cam data.

The results of the King and EFF model fitting are shown in Figs.\,7\,(a) and (b), respectively. Three concentric 1--$\sigma$ ellipses, representing the derived parameter distributions around each node, correspond to the artificial clusters of $V=17.5$, 18.5, and 19.5\,mag, when real M31 stars and variable sky background are accounted for in model fitting. We note that in both cases of the real M31 environment (crowded and with sky background gradient) the distributions of derived parameters of each node look similar, therefore, combined results of both cases are shown in Fig.\,7.

\begin{figure*}
\centering
\includegraphics[width=160mm]{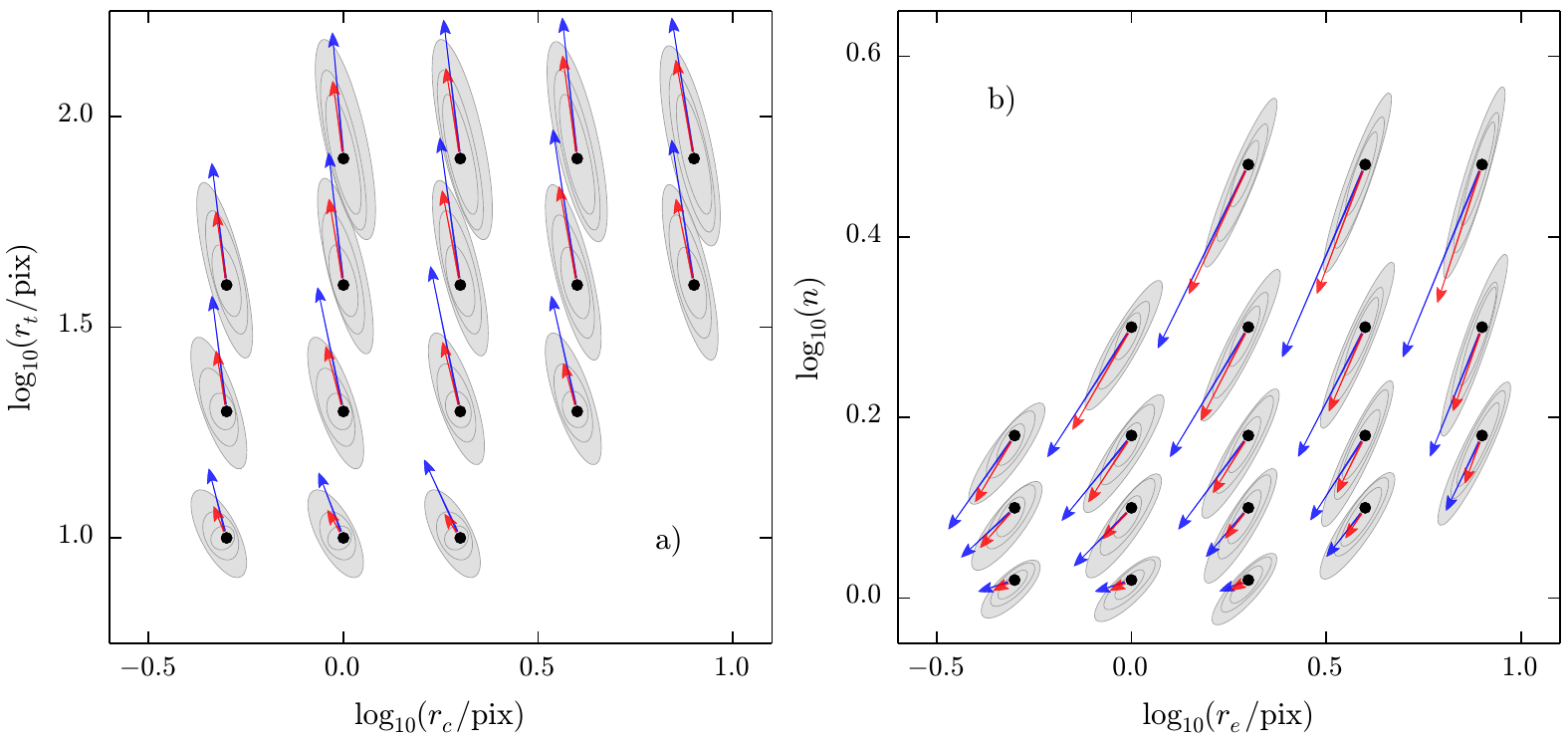}
\caption{Sky background test. Results of the King (panel a; tidal radius, $r_t$, vs core radius, $r_c$) and EFF (b; power-law index, $n$, vs scale radius, $r_e$) model fitting for artificial clusters placed in the real M31 Suprime-Cam CCD mosaic from \cite{Narbutis2008}. Black dots indicate initial parameters. For each node series of 100 frames were simulated. Three concentric 1--$\sigma$ ellipses, representing the derived parameter distributions around each node, correspond to the artificial clusters of $V=17.5$, 18.5, and 19.5\,mag. Arrows show systematic biases of derived parameter distributions, when stars and variable sky background are not taken into account in model fitting for clusters of $V=17.5$ (red/shorter) and 18.5\,mag (blue/longer).}
\end{figure*}

Arrows show systematic biases of the derived parameter distributions when stars and variable sky background are not taken into account in model fitting for clusters of $V=17.5$ (red/shorter) and 18.5\,mag (blue/longer). Significant shifts larger than 2--$\sigma$ of the distributions are observed; however, they are even more severe in the case of $V=19.5$\,mag.

Comparing derived parameters of the models presented in Fig.\,7 to the corresponding models in Figs.\,5 and 6, we note that ellipses of the derived parameter distributions (in Fig.\,7) are aligned more parallel to the vertical axes, indicating significant effect of the real sky background.

Correctly derived parameters of clusters (Fig.\,7) suggest that {\sc FitClust} can deal with ground-based images of crowded fields with variable sky background successfully for clusters with smooth profiles.

\subsection{Stochastic cluster test}
Previous tests used smooth cluster models, which are a good approximation of massive old clusters. Younger objects of lower mass are affected by stochastic luminosity fluctuations of stars and their random spatial distribution. To test performance of {\sc FitClust} on them, we followed a method described by \cite{Larsen2011} to simulate clusters star-by-star and used the SimClust program \cite{Deveikis2008} to build $V$-band images of models located in M31 observed with Suprime-Cam.

We investigated two cluster model cases: 1) 100 Myr age with mass $10^4$\,${\rm M}_\odot$, and 2) 10 Gyr age with mass $10^5$\,${\rm M}_\odot$. They represent most of the young disk clusters and old globular clusters from the study by \cite{Vansevicius2009}, and have typical fluxes of $5 \times 10^5$ and $1.3 \times 10^5$ photons in simulated images, respectively.

Stars were distributed according to the King model considering six combinations of core and tidal radius nodes with $r_c = 0.8, 1.5, 3.0$ pixels, and $r_t = 15, 40$ pixels (see Fig.\,8). One hundred cluster images were generated per node. To estimate the influence of a bright star superposed on a cluster, two additional image sets were generated by placing a star (1\,mag or 2\,mag fainter than the typical total flux of a cluster) at a distance equal to the FWHM of a cluster from its center. Uniform sky background was modeled assuming a flux of $10^3$ photons per pixel. The Gaussian photon noise was added to each pixel of the frame, assuming its standard deviation to be equal to the square root of the number of photons in a pixel.

\begin{figure*}
\centering
\includegraphics[width=160mm]{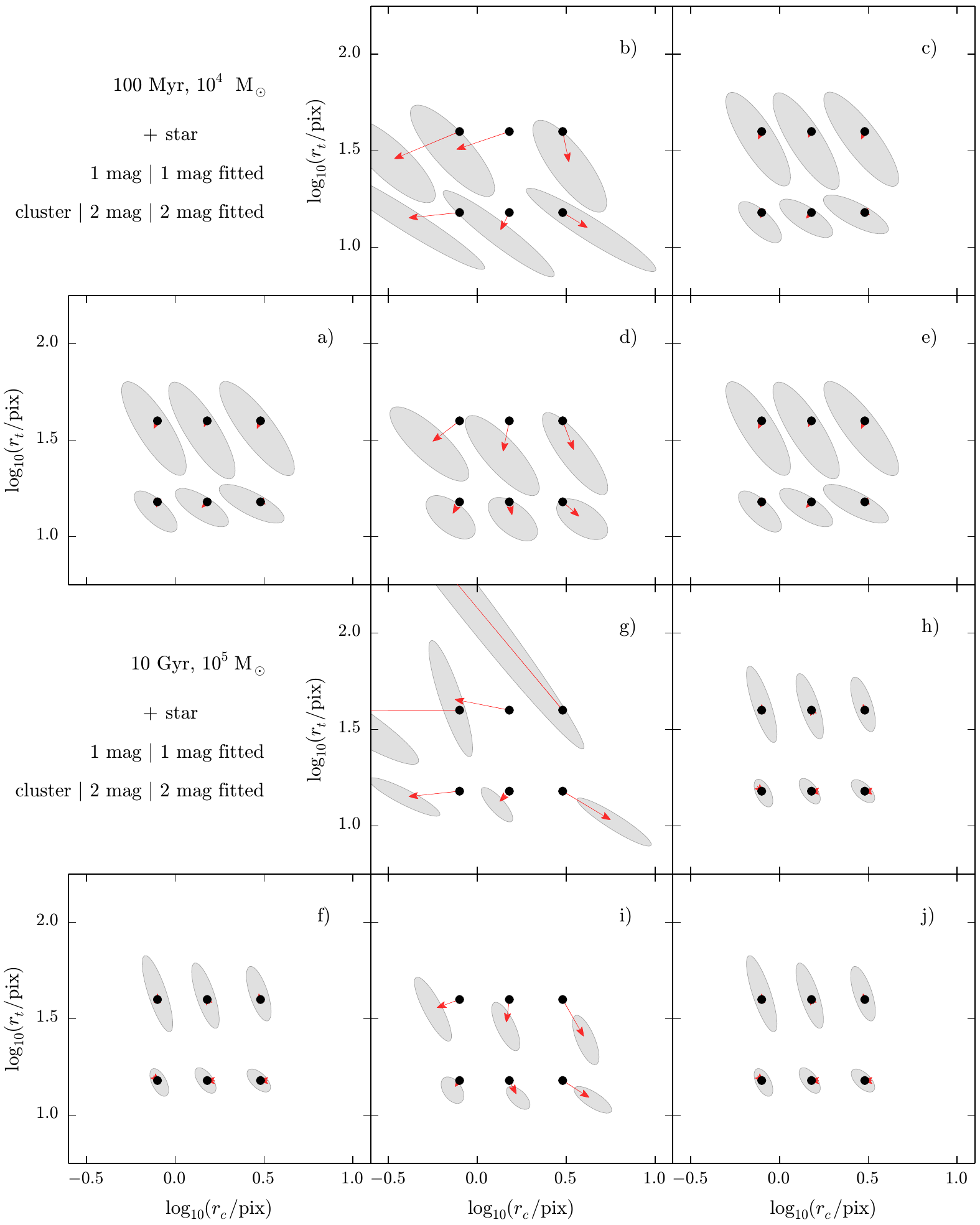}
\caption{Stochastic cluster test. Panels show King model clusters of age and mass: (a) -- (e) 100 Myr, $10^4$\,${\rm M}_\odot$, (f) -- (j) 10 Gyr, $10^5$\,${\rm M}_\odot$. In each panel dots indicate input $r_c$ and $r_t$ values of six nodes, ellipses approximate 1--$\sigma$ scatter of the derived parameters of 100 models per node, and arrows show their bias. Panels (a) and (f) show original models. Panels (b) and (g) are cases where the cluster is superposed with a star 1\,mag fainter than the cluster, (d) and (i) -- 2\,mag fainter. Star was placed at a distance equal to FWHM of cluster from its center. Panels (c), (e), (h), and (j) have the same star superposed (see corresponding row), but here the star is included in the model fit.}
\end{figure*}

Since the purpose of this test is to demonstrate influence of stochastic effects on derived structural parameters, we turned off automatic detection of stars and performed model fitting using MCMC sampling, starting from the given parameters of the simulated cluster and a bright superposed star, i.e., assuming that the position of the included star is known.

We note that the uncertainties of parameters reported by the MCMC for a single cluster are smaller than the size of dots indicating positions of grid nodes in Fig.\,8. Therefore, they are significantly underestimated and do not show true uncertainties of parameters for clusters of given age and mass.

Ellipses in Fig.\,8\,(a) approximate 1--$\sigma$ derived parameter scatter for 100 models per node of 100 Myr age, while Fig.\,8\,(f) of 10 Gyr age clusters. The derived parameter values are centered around input value. However, uncertainties due to stochastic effects, especially of tidal radius, $r_t$, are two times larger for 100\,Myr clusters than for 10\,Gyr ones and are significant.

When a bright star was included in the image, a cluster was fitted in two ways: using only the cluster model (Figs.\,8\,b,\,d,\,g,\,i), and including the star in the fit (Figs.\,8\,c,\,e,\,h,\,j). If a background star, which is 1\,mag fainter than the cluster, is superposed onto 100 Myr object (Fig.\,8\,b), significant parameter bias (shown with arrows) is observed. A similar, but weaker effect is seen when the superposed star is 2\,mag fainter than the cluster (Fig.\,8\,d).

The same effect is observed for 10\,Gyr clusters, although the largest objects in Fig.\,8\,(g) show extreme bias because the cluster's center migrates to the position of the superposed star and core radius, $r_c$, decreases drastically, while the tidal radius increases to compensate for asymmetric luminosity distribution around it. Once a star is included in the model fitting, parameters of the cluster are recovered reliably (Figs.\,8\,c,\,e,\,h,\,j); their scatter and values are close to the ones derived for cluster images without included star.

We have found that although the initial position of the input star is known, it is slightly adjusted (up to $\sim$0.2 pixel and $\sim$0.5 pixel for the 1\,mag and 2\,mag star cases, respectively) during the MCMC burn-in phase due to the underlying stochastic fluctuations on a 100\,Myr cluster. We have made tests by allowing the initial position of the star to be offset and letting for the burn-in phase to adjust it. We have found that it has to be known with an accuracy of $\sim$1 pixel for the MCMC fitting to converge to the input parameter values. Otherwise, when there are neighboring stars of comparable luminosity, i.e., in young clusters, a star migrates from its true position. Therefore, only the brightest stars can be considered in the MCMC fitting.

\subsection{Star subtraction test}
We performed a star subtraction test to quantify the influence of the cluster's brightest stars on the accuracy of its derived structural parameters. A set of cluster model images of 100\,Myr, described in Sect. 3.3, was used; 1--$\sigma$ distributions of their recovered parameters are repeated in all panels of Fig.\,9 as open ellipses centered on input parameter values.

\begin{figure*}
\centering
\includegraphics[width=160mm]{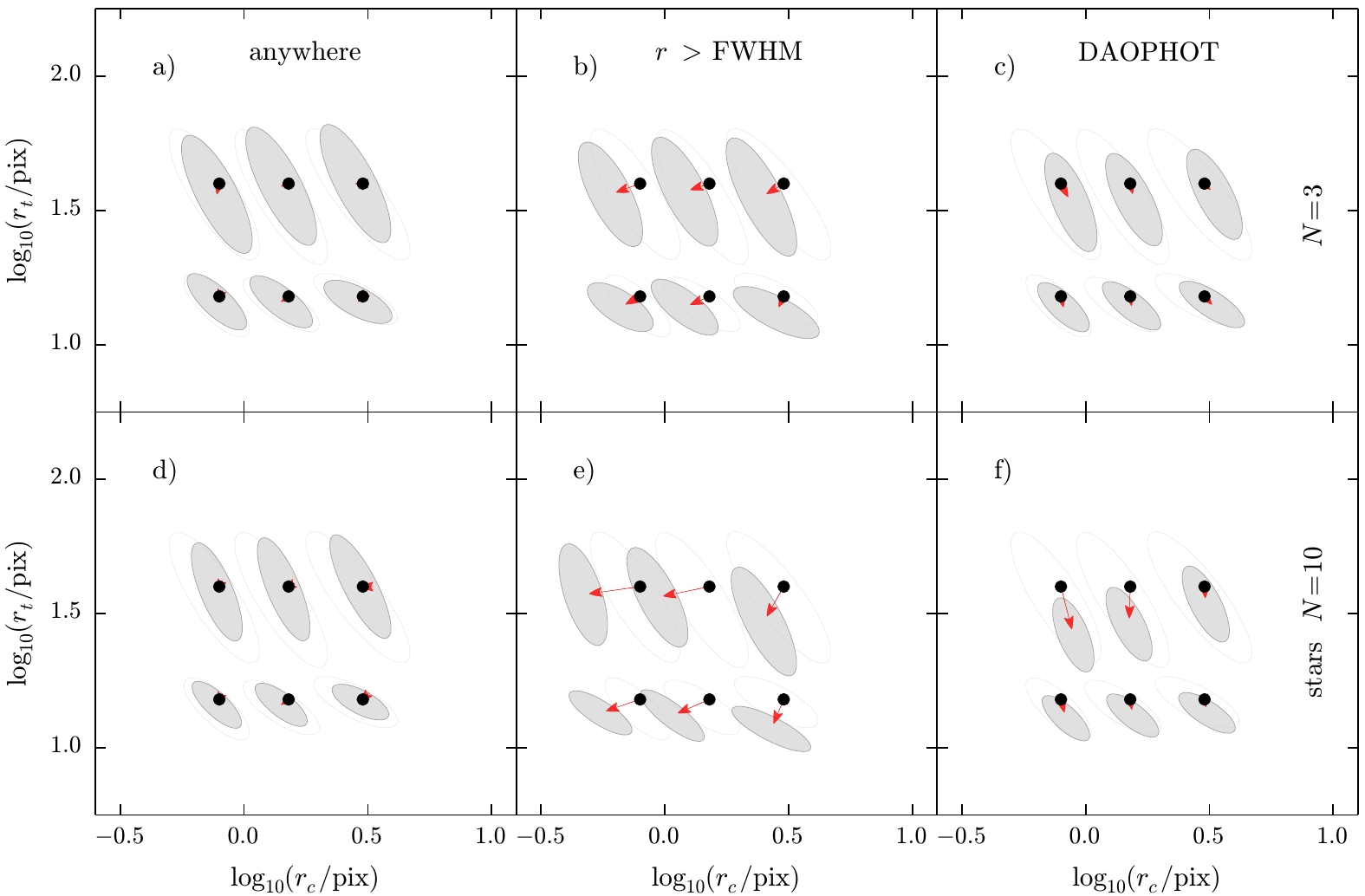}
\caption{Star subtraction test on 100\,Myr and $10^4$\,${\rm M}_\odot$ King model clusters. 100 models per input parameter node (dot) were computed and 1--$\sigma$ distributions of their recovered parameters are repeated in all panels as open ellipses centered on the nodes. Shaded ellipses indicate distributions of recovered parameters after subtraction of $N=3$ (top row) and 10 brightest stars (bottom row) of the cluster. Panels show: (a) and (d) results without these stars located anywhere in the cluster; (b) and (e) without these stars beyond cluster's FWHM, arrows indicate parameter bias; (c) and (f) the performance of {\sc FitClust} to iteratively fit model and include the 3 and 10 brightest stars detected freely with DAOPHOT.}
\end{figure*}

We have modified the original images by subtracting the three brightest stars located anywhere in the cluster (i.e., in the core or in the outskirts) and derived their parameters, which are displayed in Fig.\,9\,(a). This was repeated by subtracting the ten brightest stars; Fig.\,9\,(d) shows that parameter uncertainty has decreased two times. Therefore, if those ten brightest stars were resolved, their stochastic influence could be minimized with the MCMC model fitting. Deriving parameters of clusters without brightest stars serves as an illustration of the best cluster parameter accuracy that can be obtained.

Figures 9\,(b) and (e) show model fitting results for the same original images, but here without the three and ten brightest stars, which are located beyond the cluster's FWHM (i.e., in the outskirts), respectively. This test illustrates how parameters are affected when stars located in a cluster's outskirts are preferentially included in model fitting. The size of ellipses indicating parameter uncertainty is smaller than for clusters with all stars; however, core and tidal radii are underestimated.

Since {\sc FitClust} relies on DAOPHOT to detect stars, it is expected that stars from a cluster's outskirts would be preferentially included in a model fit, as displayed in Fig.\,2\,(c), and could result in underestimated tidal radius of the cluster. To test performance of {\sc FitClust}, we have used the same original images and performed the iterative model fitting procedure described in Sect. 2 by including the three and ten brightest stars detected freely with DAOPHOT. The results are shown in Figs.\,9\,(c) and (f). As in the previous test, the tidal radius of the cluster is underestimated, which illustrates the effect of preferential inclusion of stars in the outskirts. However, DAOPHOT identifies blended groups of stars in a cluster's core and treats them as point sources, therefore the derived core radius stays unaffected, while influence of stochasticity on parameter uncertainty is minimized. To avoid bias of the tidal radius, only a few brightest stars can be considered, as shown in Fig.\,9\,(c).

\section{Discussion and conclusions}
We presented a program, {\sc FitClust}, developed for automatic derivation of structural parameters of semi-resolved extragalactic star clusters, located in crowded stellar fields. It provides cluster structural parameters and photometry required for evolutionary studies of cluster populations in various galactic environments. The program was tested on real and simulated Subaru Suprime-Cam observations in the M31 galaxy disk.

Recently \cite{Brewer2013} attempted to identify all stars in a simulated frame of a crowded field and to derive their luminosity function by performing a catalog sampling with a variable star number. It took about one day to perform analysis of $1\,000$ stars scattered in a frame of $100 \times 100$ pixels in size on a modern multicore PC. As \cite{Brewer2013} suggested, such recovery of crowded fields could also be applicable for analysis of clusters by parameterizing spatial distribution of stars, i.e., derive their structural parameters.

{\sc FitClust} models cluster as a smooth luminosity distribution with several superposed bright stars and can derive their parameters in a few minutes on multicore PC. \cite{Narbutis2007} studied effects of aperture size on the accuracy of cluster photometry and found that field stars compromise measurements significantly. {\sc FitClust} provides magnitudes of resolved stars and the unresolved cluster component, which can be used as photometric input data to derive its evolutionary parameters, e.g., \cite{deMeulenaer2013} and \cite{Beerman2012}. Photometry of the resolved stars can be used to constrain the age of clusters, once color-magnitude diagrams are constructed and analyzed together with colors of the unresolved cluster component.

Stochastic fluctuations in a cluster's luminosity profile and sky background limit the accuracy of derived structural parameters. While in highly crowded background it is necessary to include background stars in the model fitting to recover parameters of a smooth cluster, in a semi-resolved case stars from a cluster's outskirts would be preferentially detected by DAOPHOT, resulting in underestimated cluster size. Comparing results of star subtraction test, bright star test, and sky background test, we recommend that the user of {\sc FitClust} has to set the maximum number of brightest stars to be included in model fit to avoid cluster size underestimation in semi-resolved conditions.

We stress that derived parameter uncertainties are underestimated if, e.g., only the report of MCMC fitting is used. To estimate influence of variable sky background, the image of the cluster can be co-added with a series of background subframes taken close to the cluster and each analyzed separately to build a distribution of derived parameters. Finally, the uncertainty due to stochastic effects, which is dominant for young low mass objects, can be estimated by simulating cluster observations on a star-by-star basis.

In this paper we proposed a method to account for bright field and cluster stars as well as sky background variations on the derived cluster parameters. We have demonstrated a significant influence of stochastic effects which are the main source of structural parameter uncertainty.

The strong points of the implemented algorithm are
\begin{itemize}
\item ability to derive cluster structural parameters, which are robust against background variations and bright field stars;
\item ability to perform simultaneous photometry of a cluster and resolved bright stars.
\end{itemize}

In future {\sc FitClust} versions we will optimize its computing performance and implement consistent treatment of multiband observations.
\newline
\begin{acknowledgements}
This research was funded by a grant (No. MIP-102/2011) from the Research Council of Lithuania. We thank the anonymous referee for valuable comments and the suggestion to analyze realistic images of star clusters, revealing the significance of stochasticity.
\end{acknowledgements}

\end{document}